\documentclass{ws-mpla}

\begin{document}

\def\G{\Gamma}
\def\ol{\overline}

\markboth{Yo and Nester} {Dynamic Scalar Torsion and Oscillating Universe}

%%%%%%%%%%%%%%%%%%%%% Publisher's Area please ignore %%%%%%%%%%%%%%
\catchline{}{}{}{}{}
%%%%%%%%%%%%%%%%%%%%%%%%%%%%%%%%%%%%%%%%%%%%%%%%%%%%%%%%%%%%%%%%%%%

\title{Dynamic Scalar Torsion and an Oscillating Universe}

\author{Hwei-Jang Yo}
\address{Department of Physics, National Cheng-Kung University,
Tainan 701, Taiwan, ROC\\
hjyo@phys.ncku.edu.tw}

\author{James M. Nester}
\address{Department of Physics and Institute of Astronomy, National Central University,
Chungli 320, Taiwan, ROC\\
nester@mail.phy.ncu.edu.tw}

\maketitle

%\pub{Received (Day Month Year)}{Revised (Day Month Year)}

\begin{abstract}
For the Poincar{\'e} gauge theory of gravity we consider the
dynamical scalar torsion mode in a cosmological context. We explore
in particular the possibility of using dynamical torsion to explain
the current state of the accelerating Universe. With certain
suitable sets of chosen parameters, this model can give a
(qualitatively) proper description of the current universe without
a cosmological constant, and the universe described is oscillating
with a period of the Hubble time. 
\end{abstract}

\ccode{PACS Nos.: 98.80.Cq, 98.80.Hw, 04.20.Jb, 04.50}
%%%%%%%%%%%%%%%%%%%%%%%%%%%%%%%%%%%%%%%%
\section{Introduction}
%%%%%%%%%%%%%%%%%%%%%%%%%%%%%%%%%%%%%%%%
The accelerating expansion of the Universe shown independently by
two groups\cite{Riess98,Perl99} established the existence of dark
energy with a negative pressure. The idea of a dark energy becomes
one of the greatest challenges for our current understanding of
fundamental physics.\cite{PPRB03,PadT03,CoST06} Among a number of
possibilities to describe this dark energy component, the simplest
and most theoretically appealing way may be by means of a
cosmological constant $\Lambda$, which acts on the Einstein field
equations as an isotropic and homogeneous source with a constant
equation of state $w \equiv p/\rho = -1$. Another approach toward
constructing a model for an accelerating universe is to work with
the idea that the unknown, unclumped dark energy component is due
exclusively to a minimally coupled scalar field $\Phi$ (quintessence
field) which has not yet reached its ground state and whose current
dynamics is basically determined by its potential energy $V(\Phi)$.
This idea has received much attention over the past few years and a
considerable effort  has been made in understanding the role of
quintessence fields on the dynamics of the
Universe.\cite{CaDS98,CarS98,BKBR05} However, without a specific
motivation from fundamental physics for the light scalar fields,
these quintessence models can be constructed relatively arbitrarily,
and thus there is some difficulty in relating any underlying theory
to the observable structure of the Universe.

Here we consider an accounting for the accelerated universe in terms
of an alternate gravity theory with an additional natural dynamic
geometric quantity: torsion.  Torsion has been used in several
modern theories.  The Einstein-Cartan theory has {\it non-dynamic}
torsion driven by spin density. The dominant source is spin $1/2$
fermions, which would only produce axial torsion.  The effect is
expected to be small at ordinary densities, but it could have a
major influence at high densities (e.g. beyond $10^{48}$
gm/cm${}^3$), and thus it is expected to have important physical
effects only in the early universe. Torsion cosmology investigations
were initiated by Kopc{\'n}yski.\cite{Kop}  Some early
investigations noted that torsion could prevent singularities, but
soon it was argued that {\it non-linear} torsion effects were more
likely to produce stronger singularities.\cite{IN77}

The Poincar\'e gauge theory of gravity (PGT),\cite{Hehl80,Hayash80}
based on a Riemann-Cartan geometry, allows for dynamic torsion in
addition to curvature. Because of its gauge structure and geometric
properties it was regarded as an attractive alternative to general
relativity. The various possible combinations of well behaved
propagating modes in the linear theory were identified.  Then it was
noted that effects due to the non-linear constraints could be
expected to render most of these cases physically
unacceptable.\cite{CNY98}  An investigation showed that two
particular ``scalar torsion'' cases had a well posed initial value
problem.\cite{HNZ96} Later Hamiltonian
investigations\cite{YHNJ99,YN02} supported the conclusion that these
two dynamic scalar torsion modes may well be the only acceptable
dynamic PGT torsion modes.

In one scalar torsion mode only the axial vector torsion is dynamic
(and it turns out to be dual to the gradient of a scalar field).
Axial torsion is naturally driven by the intrinsic spin of
fundamental fermions; in turn it naturally interacts with such
sources. Thus for this mode one has some observational constraints.
Note that except in the early universe one does not expect large
spin densities. Consequently it is generally thought that axial
torsion must be small and have small effects at the present time.
This is one reason why we do not focus on this mode here.

The other scalar torsion mode is vector torsion (which turns out to
depend on the gradient of a scalar field).  There is no known
fundamental source which directly excites this mode.  Conversely
this type of torsion does not interact in any direct dramatic
fashion with any familiar type of matter.   Hence we do not have
much in the way of constraints as to its magnitude.  We could
imagine it as having significant magnitude and yet not being
dramatically noticed except indirectly through the non-linear
equations.  This mode in particular has also attracted our interest
because of a conspicuous consequence of the non-linear equations: in
this case there is a critical non-zero value for the affine scalar
curvature.

 A comprehensive survey of the PGT torsion cosmological models was
presented some time ago by Goenner and M{\"u}ller-Hoissen.\cite{GMH}
Although this work only solved in detail a few particular cases, it
developed the equations for all the PGT cases, including those for
the particular model we consider here. Since this work was done
prior to the discovery of the accelerating universe, they generally
imagined torsion as perhaps playing a big role only at high
densities in the early universe. More recently investigators have
begun to consider torsion as a possible cause of the accelerating
universe (see e.g.~\cite{Boe03,CaCT03}) but the subject has not yet
been explored in detail.

Our theoretical PGT analysis led us to consider just two dynamic
torsion modes. An obvious place where we might see some physical
evidence for these modes is in cosmological models.  The homogeneous
and isotropic assumptions of cosmology greatly restrict the possible
types of non-vanishing fields.  Curiously, for torsion there are
only two possibilities: vector torsion which, moreover, is the
gradient of a function (which depends only on time) and axial
torsion which is dual to the gradient of a function (depending only
on time).  Thus the homogeneous and isotropic cosmologies are
naturally very suitable for the  exploration of the physics of the
dynamic PGT scalar modes.

In this work, we take our first step to explore the possible
evolution of the Universe with the scalar torsion mode of PGT. The
main motivation is two-fold: (1) to have a better understanding of
the PGT, in particular the possible physics of the dynamic scalar
torsion modes; (2) to consider the prospects of accounting for the
outstanding present day mystery---the accelerating universe---in
terms of an alternate gravity theory, more particularly in terms of
the PGT dynamic torsion. With the usual assumptions of isotropy and
homogeneity in cosmology, we find that, under the model, the
Universe will oscillate with generic choices of the parameters. The
torsion field in the model plays the role of the imperceptible
``dark energy''. With a certain range of parameter choices, it can
account for the  current status of the Universe, i.e., an
accelerating expanding universe with a value of the Hubble constant
which is approximately the present one. These promising results
should encourage further investigations of this model, with a
detailed comparison of its predictions with the observational data.

This report is organized as follows: We summarize the formulation of
the scalar torsion mode of PGT in Sec. II and the expressions of
this model in cosmology in Sec. III. We give a preliminary
analytical analysis of the behavior of the field equations of this
model in Sec. IV. In Section V we present the results of our
numerical demonstrations for various choices of the parameters and
the initial data. We discuss the implications of our findings in
Sec. VI.
%%%%%%%%%%%%%%%%%%%%%%%%%%%%%%%%%%%%%%%%
\section{Torsion field equation}
%%%%%%%%%%%%%%%%%%%%%%%%%%%%%%%%%%%%%%%%%
The affine connection of the Riemann-Cartan geometry has the form
\begin{equation}
   {\G}^k{}_{ij} = \ol{\G}^k{}_{ij}+\frac{1}{2}(T_{ji}{}^k
               + T^k{}_{ij} + T^k{}_{ji}),
\end{equation}
where $\ol{\G}^k{}_{ij}$ is the Levi-Civita connection and
$T_{ij}{}^k$ is the torsion tensor. The Ricci curvature and scalar
curvature of the Riemann-Cartan geometry can be expanded in the form
\begin{eqnarray}
   R_{ij} &=& \ol{R}_{ij} + \ol{\nabla}_jT_i+\frac{1}{2}(\ol{\nabla}_k - T_k)
   (T_{ji}{}^k+T^k{}_{ij}+T^k{}_{ji})\nonumber\\&&
   +\frac{1}{4}(T_{kmi}T^{km}{}_j+2T_{jkm}T^{mk}{}_i),\\
   R&=&\ol{R} + 2\ol{\nabla}_iT^i+\frac{1}{4}(T_{ijk}T^{ijk}
    +2T_{ijk}T^{kji}-4T_iT^i),
\end{eqnarray}
where $\ol{R}_{ij}$ and $\ol{R}$ are the Riemannian Ricci curvature
and scalar curvature, and $\ol\nabla$ is the covariant derivative with
the Levi-Civita connection.

For the case of interest here the torsion tensor has the restricted
form
\begin{equation}
   T_{ijk} = \frac{2}{3}T_{[i}g_{j]k},
\end{equation}
where the vector $T_i$ is the trace of the torsion. Consequently the
affine Ricci curvature and scalar curvature become
\begin{eqnarray}
   R_{ij}&=&\ol{R}_{ij} + \frac{1}{3}(2\ol{\nabla}_jT_i+g_{ij}\ol{\nabla}_kT^k)
   +\frac{2}{9}(T_iT_j-g_{ij}T_kT^k),\\
   R&=&\ol{R} + 2\ol{\nabla}_iT^i-\frac{2}{3}T_iT^i.\label{aRvsrR}
\end{eqnarray}
For the PGT scalar torsion mode the gravitational Lagrangian density
is\cite{YHNJ99} %([\refcite{YHNJ99}])
\begin{equation}
  L = - \frac{a_0}{2}R + \frac{b}{24}R^2+ \frac{a_1}{8}\left(
            T_{ijk}T^{ijk}+2T_{ijk}T^{kji}-4T_i T^i\right).
\end{equation}
Then the field equations of concern are
\begin{eqnarray}\label{fe1}
   &&a_0\left(\ol{R}_{ij}-\frac{1}{2}g_{ij}\ol{R}\right)-\frac{b}{6}R
        \left(R_{(ij)}-\frac{1}{4}g_{ij}R\right)\nonumber\\
    &&\qquad\qquad+\frac{2m}{3}(\ol{\nabla}_{(i}T_{j)}-g_{ij}\ol{\nabla}_kT^k)
    +\frac{m}{9}(2T_iT_j+g_{ij}T_kT^k) ={\cal T}_{ij},\\
   &&\ol{\nabla}_iR + \frac{2}{3}\left(R-\frac{6m}{b}\right)T_i=0,\label{gradR}
\end{eqnarray}
with the appropriate assumption of the source spin density vanishing
in Eq.~(\ref{gradR}). Contracting Eq.~(\ref{fe1}) using
Eq.~(\ref{aRvsrR}) gives $a_1\ol{R}+mR+{\cal T}=0$. Equation
(\ref{fe1}) can now be re-written in an effective Einstein equation
form:
\begin{equation}
   a_0\left(\ol{R}_{ij}-\frac{1}{2}g_{ij}\ol{R}\right)=\tau_{ij}:={\cal T}_{ij}
   +\ol{\cal T}_{ij},\label{effeineq}
\end{equation}
where the effective stress-energy tensor due to the scalar torsion
is given by
\begin{eqnarray}\!\!
   \ol{\cal T}_{ij}&=&\frac{b}{6}R
        (R_{(ij)}-\frac{1}{4}g_{ij}R)
    -\frac{2m}{3}(\ol{\nabla}_{(i}T_{j)}-g_{ij}\ol{\nabla}_kT^k)
    -\frac{m}{9}(2T_iT_j+g_{ij}T_kT^k).\label{torT}
\end{eqnarray}
Relation~(\ref{effeineq}) is quite useful both intuitively and
practically, even though it has the shortcoming that all the second
derivatives of the metric have not been isolated on one side of the
equation.
%%%%%%%%%%%%%%%%%%%%%%%%%%%%%%%%%%%%%%%%
\section{Cosmology}
%%%%%%%%%%%%%%%%%%%%%%%%%%%%%%%%%%%%%%%%
With the cosmological assumption of isotropy and homogeneity, the
spacetime metric takes on the FRW form,
\begin{equation}
   ds^2 = - dt^2+a^2(t)\left[\frac{dr^2}{1-kr^2}+r^2(d\theta^2+\sin^2\theta
          d\phi^2)\right];
\end{equation}
here we consider only the simplest case, the flat universe, $k=0$.
Then the nonvanishing Riemannian Ricci and scalar curvature are the
well known
\begin{equation}
   \ol{R}_t{}^t = 3(\dot{H} + H^2),\quad
   \ol{R}_r{}^r = \ol{R}_\theta{}^\theta = \ol{R}_\phi{}^\phi
                = \dot{H} + 3H^2,\quad
   \ol{R} = 6(\dot{H} + 2H^2),\label{trR}
\end{equation}
where $H$ is the Hubble rate $H=\partial_t(\ln a)$.

In accordance with the homogeneity and isotropy the torsion vector
$T_i$ depends only on time, and has just one non-vanishing
component: $\Phi(t):=T_t(t)$.
Using (\ref{trR}) in the Field equations
(\ref{fe1},\ref{gradR}) we can finally give the necessary equations
to integrate:
\begin{eqnarray}
      \dot{a}&=&aH,\label{dta}\\
      \dot{H}&=&-\frac{m}{6a_1}R-\frac{1}{6a_1}{\cal T}-2H^2,\label{dtH}\\
      \dot{\Phi}&=&-\frac{a_0}{2a_1}R-\frac{1}{2a_1}{\cal T}-3H\Phi
                   +\frac{1}{3}\Phi^2,\label{dtphi}\\
      \dot{R}&=&-\frac{2}{3}\left(R-\frac{6m}{b}\right)\Phi,\label{dtR}
\end{eqnarray}
where
\begin{eqnarray}
   &&3a_1H^2+\frac{b}{24}R^2-\frac{b}{18}\left(R-\frac{6m}{b}\right)
   (3H-\Phi)^2   ={\cal T}_{tt},\label{fieldrho}\\
   &&{\cal T}^i{}_j = \hbox{diag}(-\rho,p,p,p),\quad
   {\cal T} = g^{ij}{\cal T}_{ij} = 3p-\rho,\label{trT}\\
   &&p=w\rho.
\end{eqnarray}
For a matter-dominated era, the pressure $p$ is negligible, i.e.,
$\omega\approx 0$. For the stress-energy tensor from the scalar
torsion the explicit expression is
\begin{eqnarray}
   &&\ol{\cal T}_t{}^t=\frac{b}{24}R^2-3mH^2-\frac{b}{18}\left(R
     -\frac{6m}{b}\right)(3H-\Phi)^2,\label{Trho}\\
   &&\ol{\cal T}_r{}^r=\ol{\cal T}_\theta{}^\theta=\ol{\cal T}_\phi{}^\phi
     =\frac{1}{3}[m(R-\ol{R})-\ol{\cal T}_t{}^t],\label{Tpre}
\end{eqnarray}
and the off-diagonal terms vanish.
%%%%%%%%%%%%%%%%%%%%%%%%%%%%%%%%%%%%%%%%%%%%%%%%%%%%%%
\section{A preliminary analysis of the equations}
%%%%%%%%%%%%%%%%%%%%%%%%%%%%%%%%%%%%%%%%%%%%%%%%%%%%%%
Equations (\ref{dta}--\ref{dtR}) are the main equations for the
integrations to evolve the system.  Regarding the parameters in the
field equations, the Newtonian limit requires $a_0\equiv(8\pi
G)^{-1}$. We take $a_1>0$ and $b>0$ to satisfy the energy positivity
requirement.\cite{YHNJ99} Furthermore, we assume $m>0$ since $m$ is
the mass parameter of the dynamic torsion mode.

Before the detailed results are shown, we briefly analyze the
equations, to obtain some insight about their behavior. Let us first
study the behavior of the affine scalar curvature $R$. For
convenience, we define $Y(t)\equiv R(t)-6m/b$, then we can rewrite
Eq.~(\ref{dtR}) as
\begin{equation}
  \dot{Y}=-\frac{2}{3}Y\Phi.\label{dtY}
\end{equation}
The second derivative of $Y$ with respect to time can be obtained by
operating a time derivative on Eq.~(\ref{dtY}) and using
Eq.~(\ref{dtphi}).
From this analysis we find that the
late-time behavior will be essentially affected by the initial value
$Y(t_0)$.
%Since all the coefficients of $Y$ on the rhs
%of Eq.~(\ref{ddtYa}) are positive definite
$Y$ would grow unboundedly if its initial value is chosen to be
positive. In order to have a bounded value of $Y(t)$ at late time,
the initial value of $Y$ is taken to be $Y(t_0)\le0$. Then it can be
shown from Eq.~(\ref{dtY}) %and (\ref{ddtYa})
that $Y(t)$ will always
be $Y(t)\leq0$.
%[Note that this statement is an exact result from
%(\ref{ddtYa}) instead being approximated from
% Eq.~(\ref{ddtYa}).]
 Here we will restrict ourselves to the
situation where
\begin{equation}
Y\leq0\quad\Longrightarrow\quad R\leq\frac{6m}{b},\label{Rrange}
\end{equation}
so that this model could be suitable for describing the current
status of the Universe.

Now let us turn our attention to $\Phi$, the nonvanishing part of
the torsion.  An analysis of the approximate equation of motion
leads to the conclusion that we can expect periodic behavior, with a
period around $T=2\pi\sqrt{b/2m}$ and an estimated maximal amplitude
With the same kind of  reasoning, we can find that $R$ has a similar
periodic behavior.

However there seems to be no  way to figure out the behavior of the
expansion factor $a$ with a similar analysis. The acceleration of
the expansion factor $a$ can be derived by taking a further
derivative on Eq.~(\ref{dta}) and using Eq.~(\ref{dtH}):
\begin{equation}
\ddot a=-\frac{mR+{\cal T}}{6a_1}a-\frac{{\dot a}^2}{a}.
\end{equation}
We can see from this equation that there does not exist an obvious
leading term. Therefore it is not easy to understand the behavior of
$a$ with a suitable approximation. However, the period of $a$ and
$H$, if it exists, should be the same as $\Phi$ and $R$, since the
variables are all highly coupled to each other, and thus there
should exist a common period in the solution. We demonstrate this
point with the numerical analysis in the next section.

However, we need to look into the scaling features of this model
before we can obtain the sort of results we seek on a cosmological
scale. In terms of fundamental units we can scale the variables as
\begin{equation}
t\rightarrow t/\ell_p,\,\, a\rightarrow a,\,\, H\rightarrow\ell_pH,
\,\, \Phi\rightarrow\ell_p\Phi,\,\, R\rightarrow\ell^2_pR,
\end{equation}
and the corresponding scaling of the parameters are
\begin{equation}\label{scale1}
a_0\rightarrow\ell^2_pa_0,\,\, a_1\rightarrow
\ell^2_pa_1,\,\, m\rightarrow\ell^2_pm,\,\, b\rightarrow b,
\end{equation}
where $\ell_p\equiv\sqrt{8\pi G}$. In such a case, all the
variables, and the scaled parameters $a_0$, $a_1$, and $b$, become
dimensionless, and $a_0=1$. Furthermore, equations
(\ref{dta}-\ref{dtR}) remain unchanged under such a scaling. However,
as we are interested in the cosmological scale, it is practical to
use another scaling to turn the numerical values of the scaled
variables ``gentler'' (i.e, not stiff) from the numerical
integration. In order to achieve this goal, let us introduce a
dimensionless constant $T_0$, which has the order of magnitude of
the Hubble time. Then the scaling is
\begin{eqnarray}
&&t\rightarrow T_0t,\,\, a\rightarrow a,\,\, H\rightarrow H/T_0,\,\,
\Phi\rightarrow\Phi/T_0,\,\, R\rightarrow R/T_0^2,\nonumber\\
&&a_0\rightarrow a_0,\,\, a_1\rightarrow a_1,\,\, m\rightarrow m,\,\,
b\rightarrow H^2_0 b,\label{scale2}
\end{eqnarray}
With this scaling, all the field equations are kept unchanged while
the period $T\rightarrow T_0T$.

%%%%%%%%%%%%%%%%%%%%%%%%%%%%%%%%%%%%%%%%
\section{Numerical demonstration}
%%%%%%%%%%%%%%%%%%%%%%%%%%%%%%%%%%%%%%%%
In this section we would like to demonstrate two points:  (i) In the
degenerate case, i.e., $Y=0$, the torsion in the system becomes
kinetic instead of being dynamic, and the expansion is always
decelerating; (2) In generic cases, i.e., $Y<0$, the torsion in the
system is dynamic, and the functional pattern of the expansion
factor has a periodic feature, i.e., it could be accelerating for a
while, and then be followed by a period of deceleration with the
pattern repeating. With suitable choices of the parameters and the
initial values of the fields involved, it is possible to change the
period of the dynamic fields as well as their amplitudes; (3) In the
model, with some choices of the parameters and the initial values of
the fields, it is possible to mimic the main apparent dynamic
features of the Universe, i.e., the value of the Hubble function is
the current Hubble constant in an accelerating universe after a
period of time on the order of the Hubble time. In such a case, this
model will describe an oscillating universe with a period on the
order of magnitude of the Hubble time. This allows us to constrain
the parameters and/or the value of the torsion field by comparing
the observed data with the result from this model.

\begin{figure*}[thbp]
\begin{tabular}{rl}
\includegraphics[width=6cm]{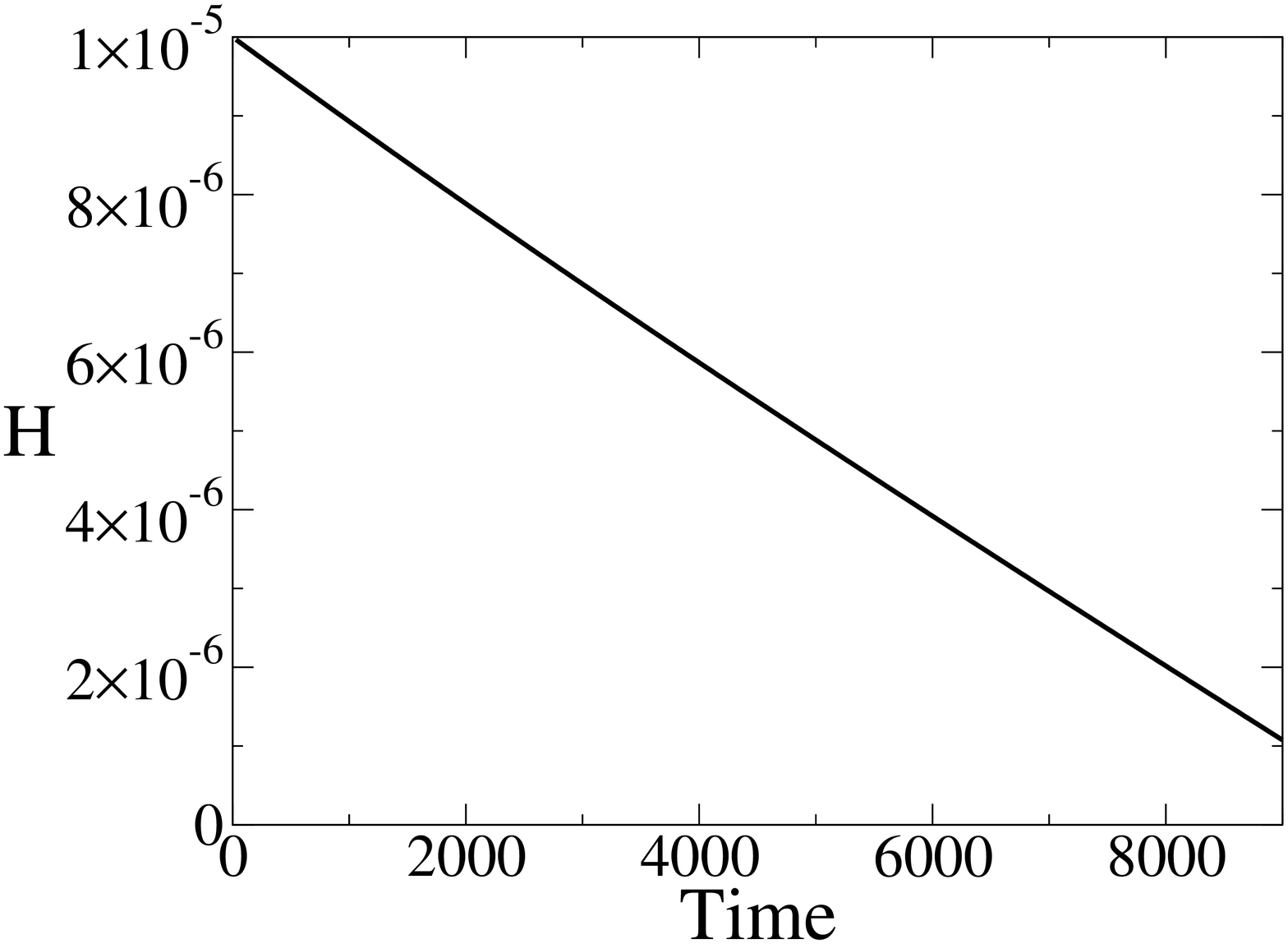}&
\includegraphics[width=6cm]{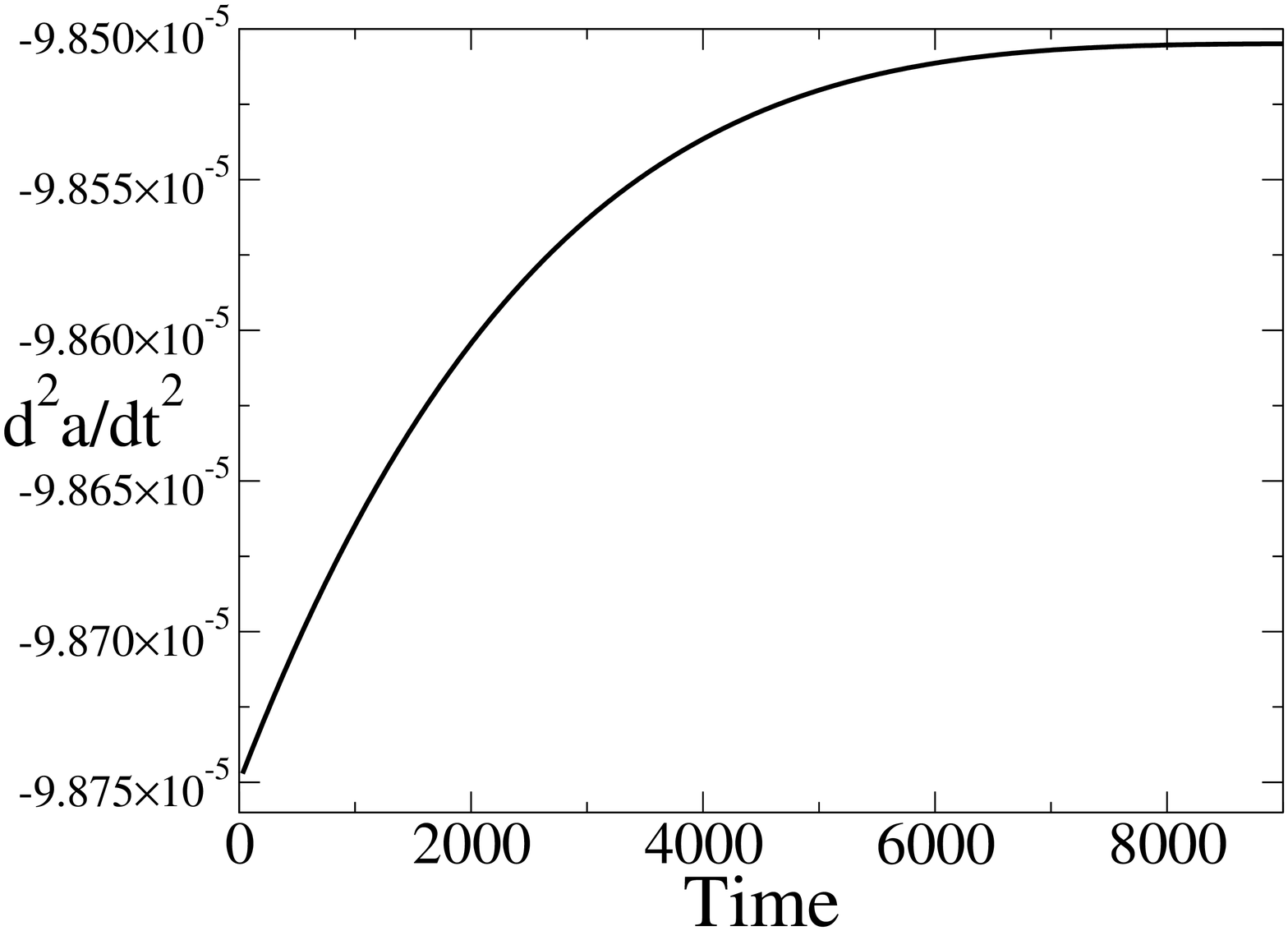} \\
\includegraphics[width=6cm]{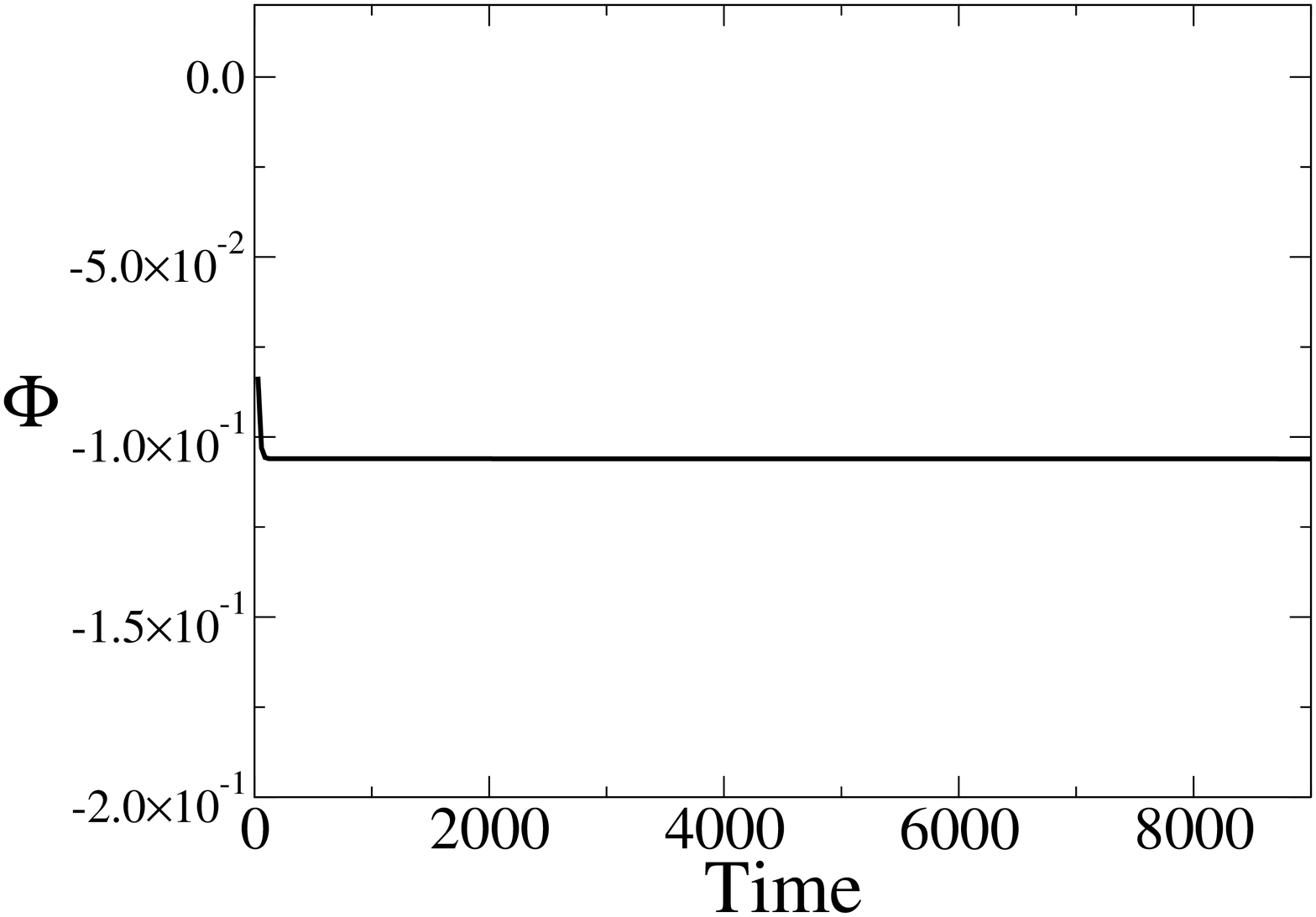}&
\includegraphics[width=6cm]{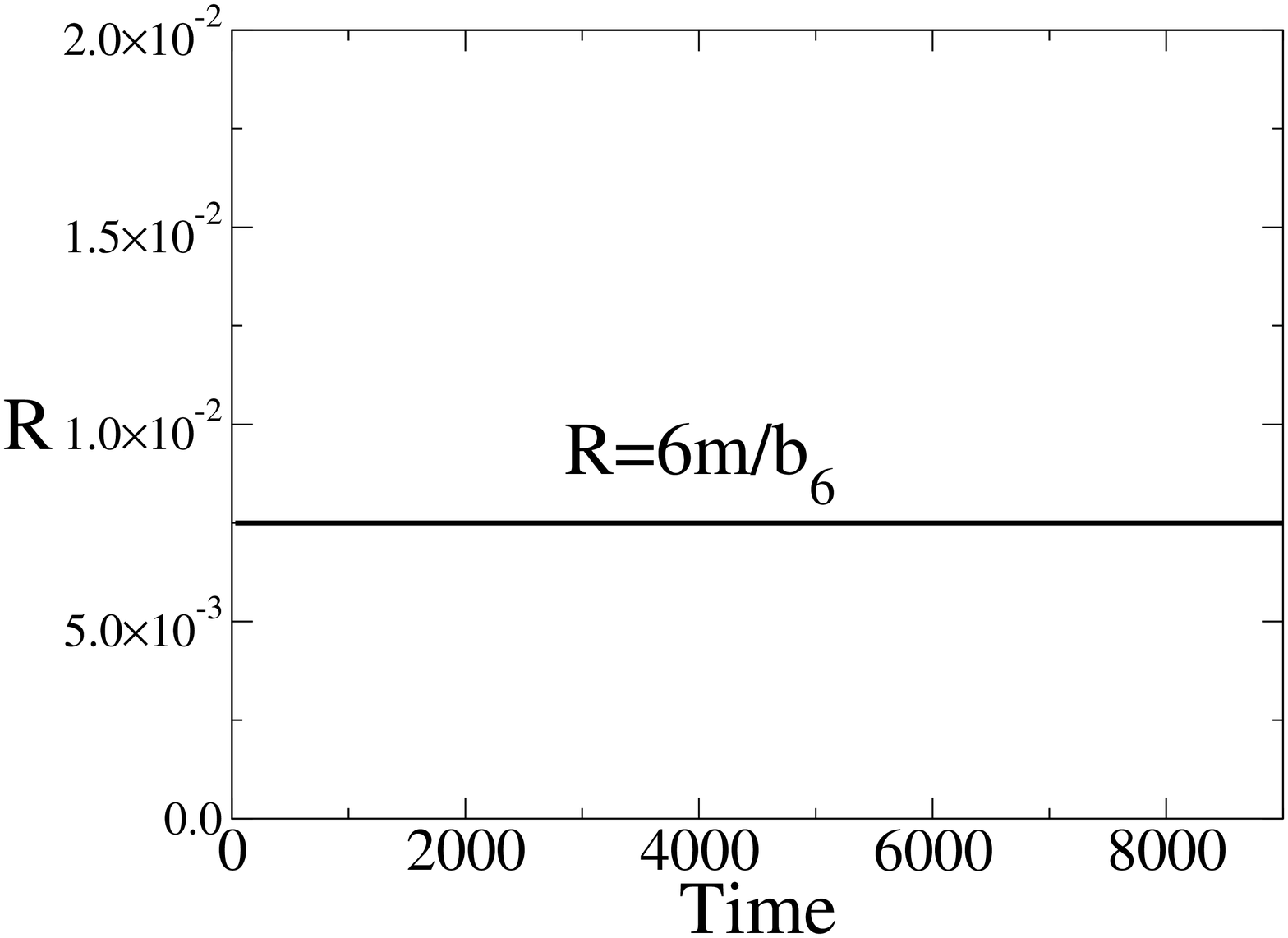}
\end{tabular}
\caption{Evolution of the Hubble function,  the 2nd time derivative
of the expansion factor, the temporal component of the torsion, and
the affine scalar curvature as functions of time with the parameter
choice and the initial data in Case I.} \label{conR}
\end{figure*}

The 4th-order Runge-Kutta method is applied for the integration of
the field equations (\ref{dta}-\ref{dtR}). The Universe is assumed
to be matter-dominated, thus ${\cal T}\approx-\rho$. The mass
density $\rho$ is determined from the fields via
Eq.~(\ref{fieldrho}). For all cases, the fields and the parameters
are scaled with Eq.~(\ref{scale1}) to be dimensionless. For the
third case, the fields and the parameters are scaled further with
Eq.~(\ref{scale2}) to achieve a realistic cosmology. In all cases we
choose $Y(t_0)\leq0$. The system will keep $Y(t)\leq0$ as long as
the initial value of $Y$ is so chosen.

\begin{figure*}[thbp]
\begin{tabular}{rl}
\includegraphics[width=6cm]{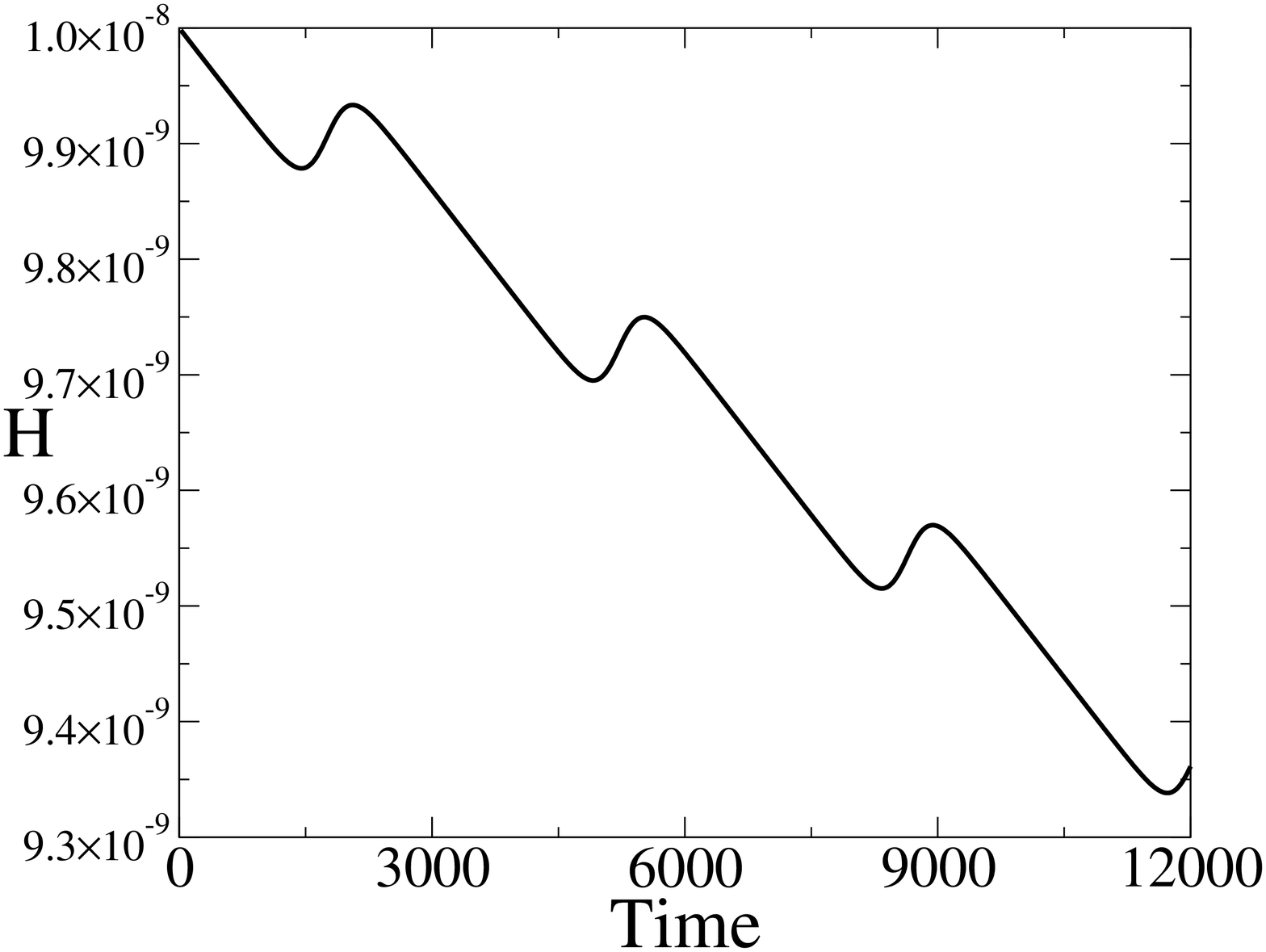}&
\includegraphics[width=6cm]{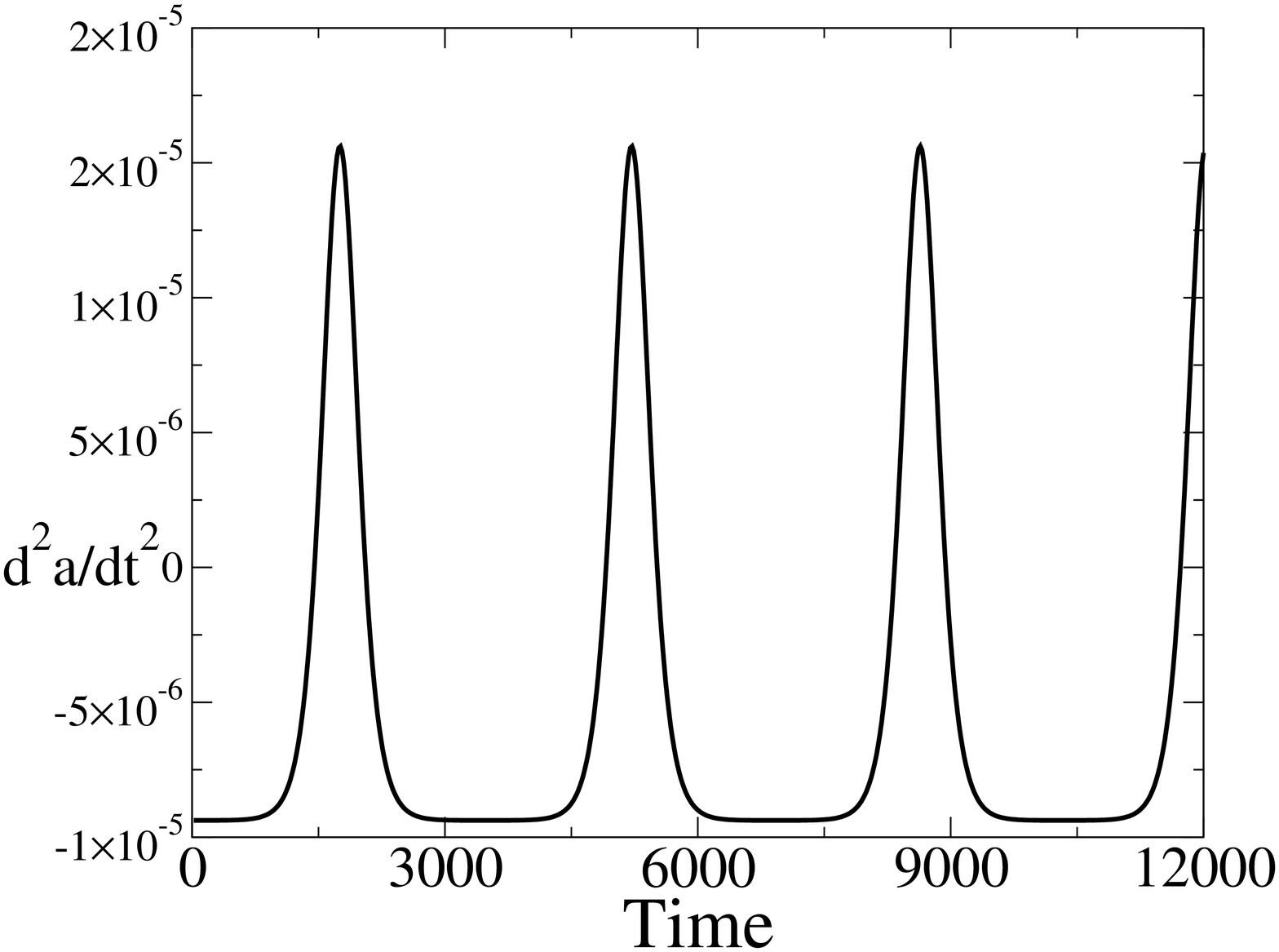} \\
\includegraphics[width=6cm]{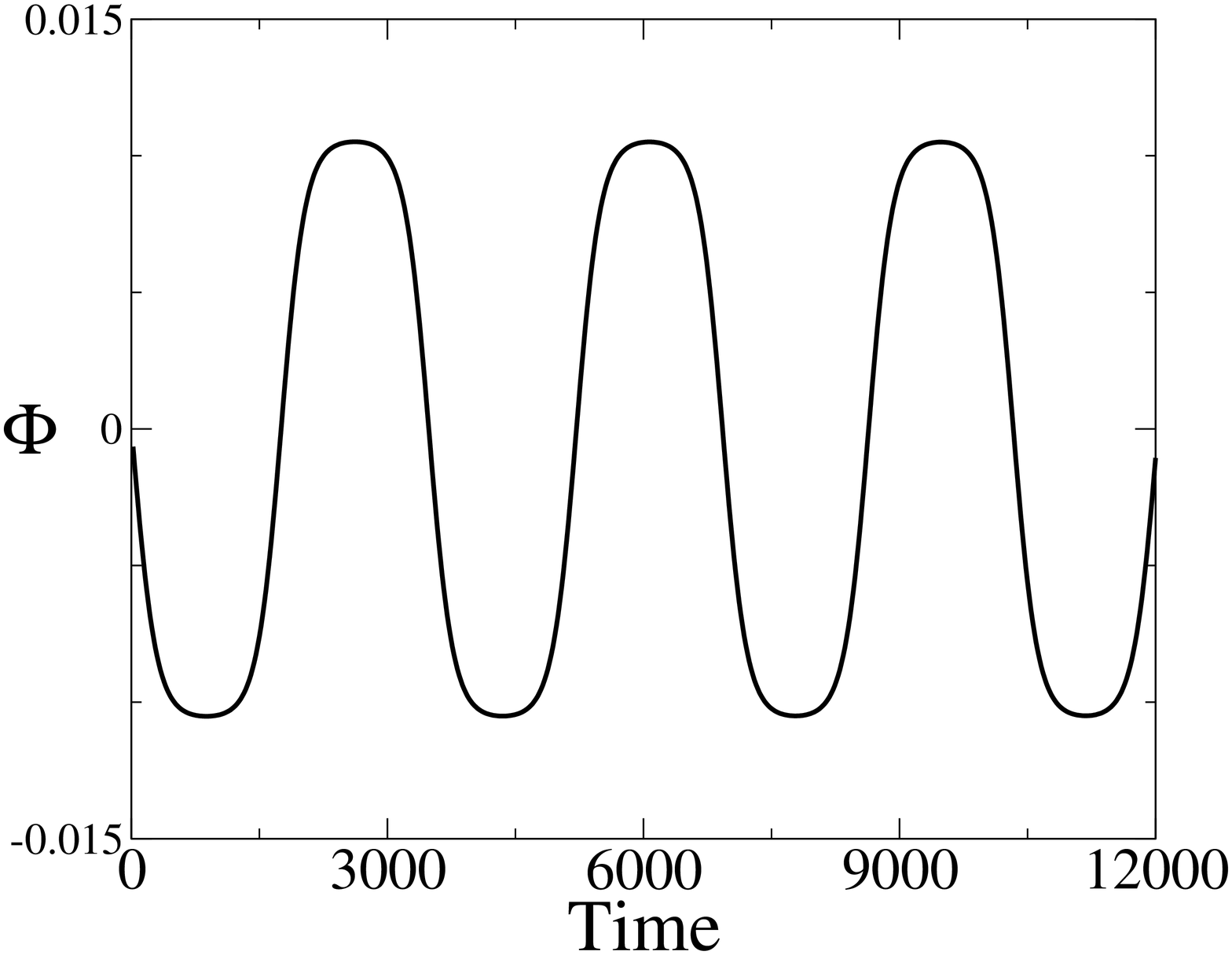}&
\includegraphics[width=6cm]{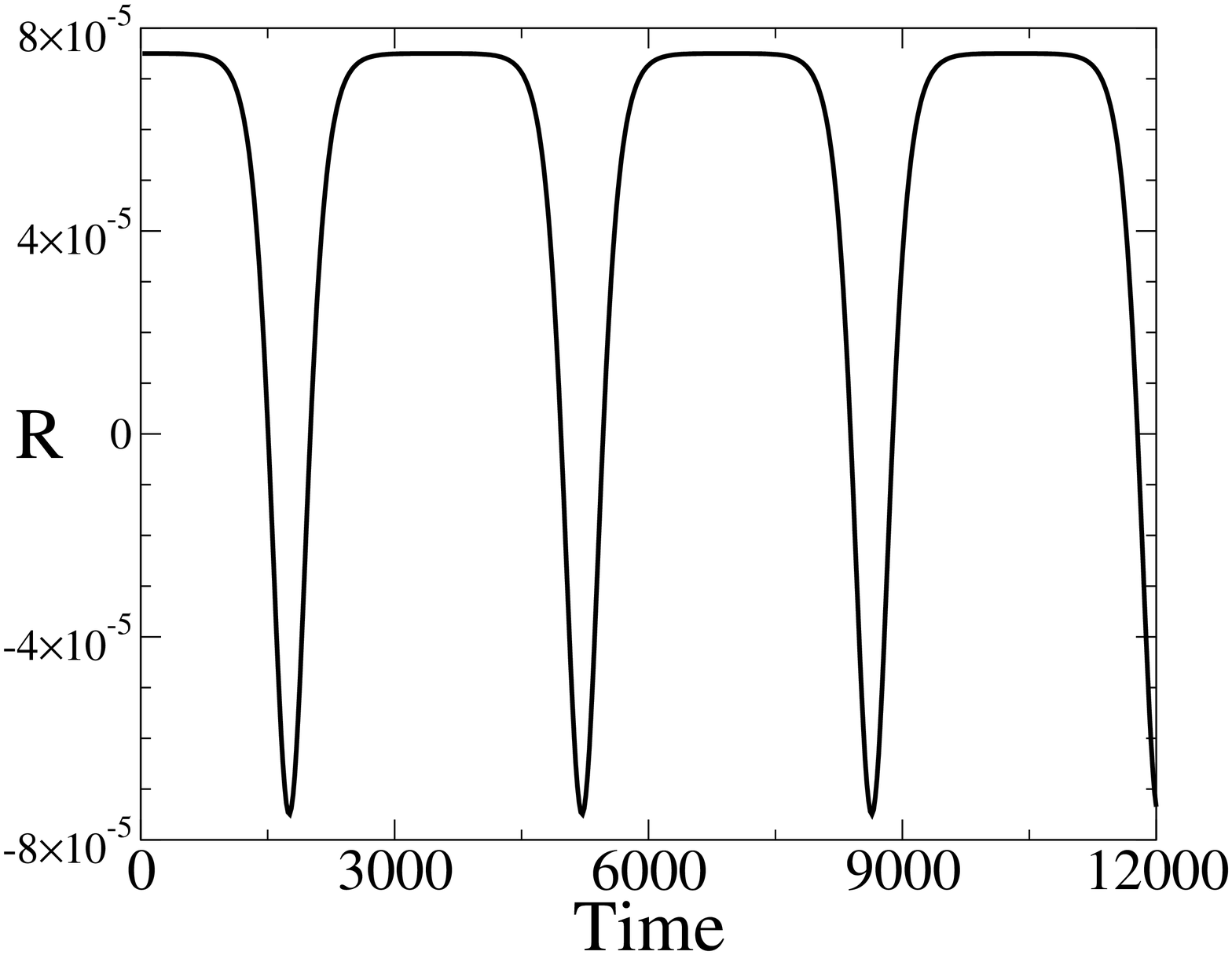}
\end{tabular}
\caption{Evolution of the Hubble function, the 2nd time derivative
of the expansion factor, the temporal component of the torsion, and
the affine scalar curvature as functions of time with the parameter
choice and the initial data in Case II.} \label{oscR1}
\end{figure*}

\begin{figure*}[thbp]
\begin{tabular}{rl}
\includegraphics[width=6cm]{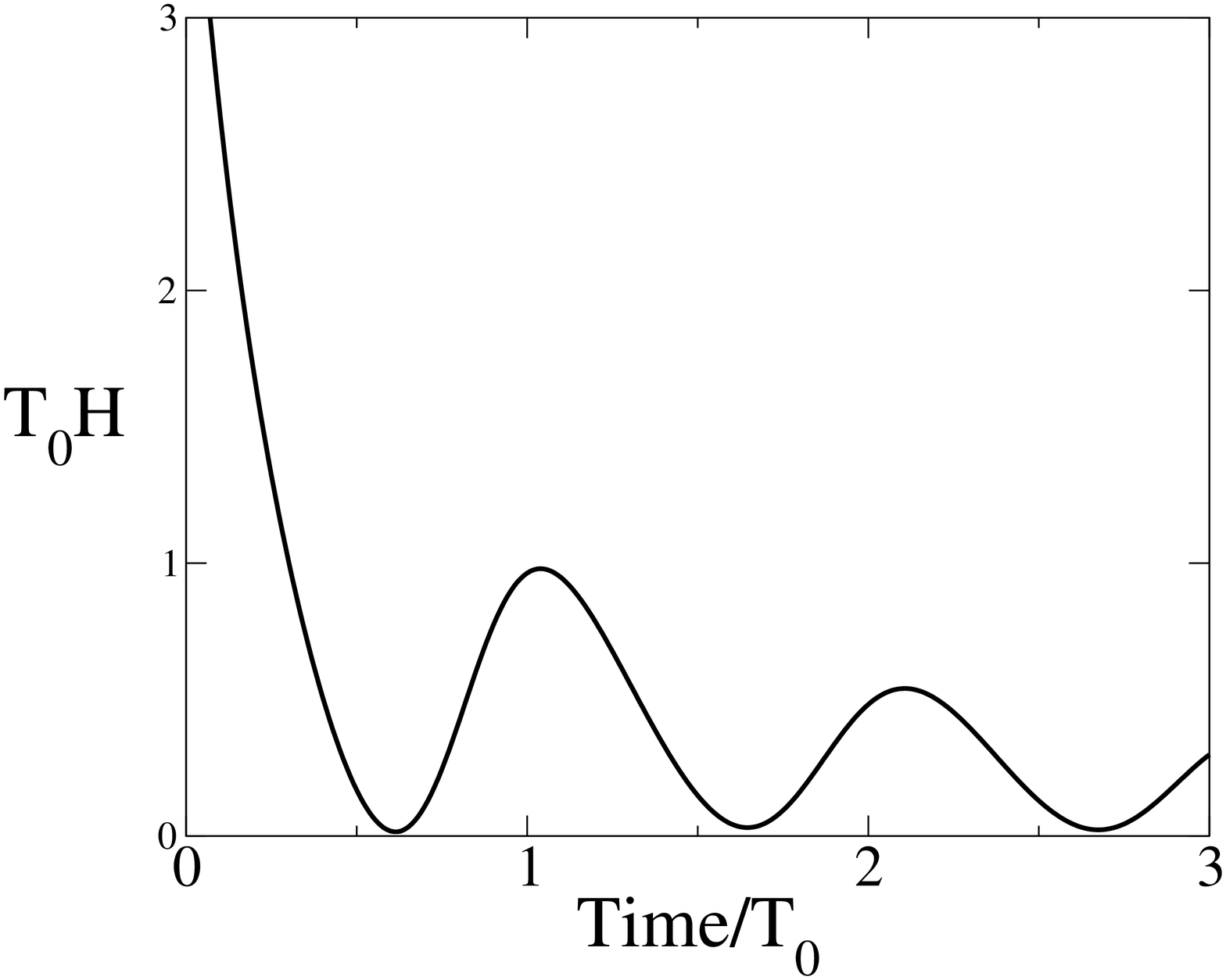}&
\includegraphics[width=6cm]{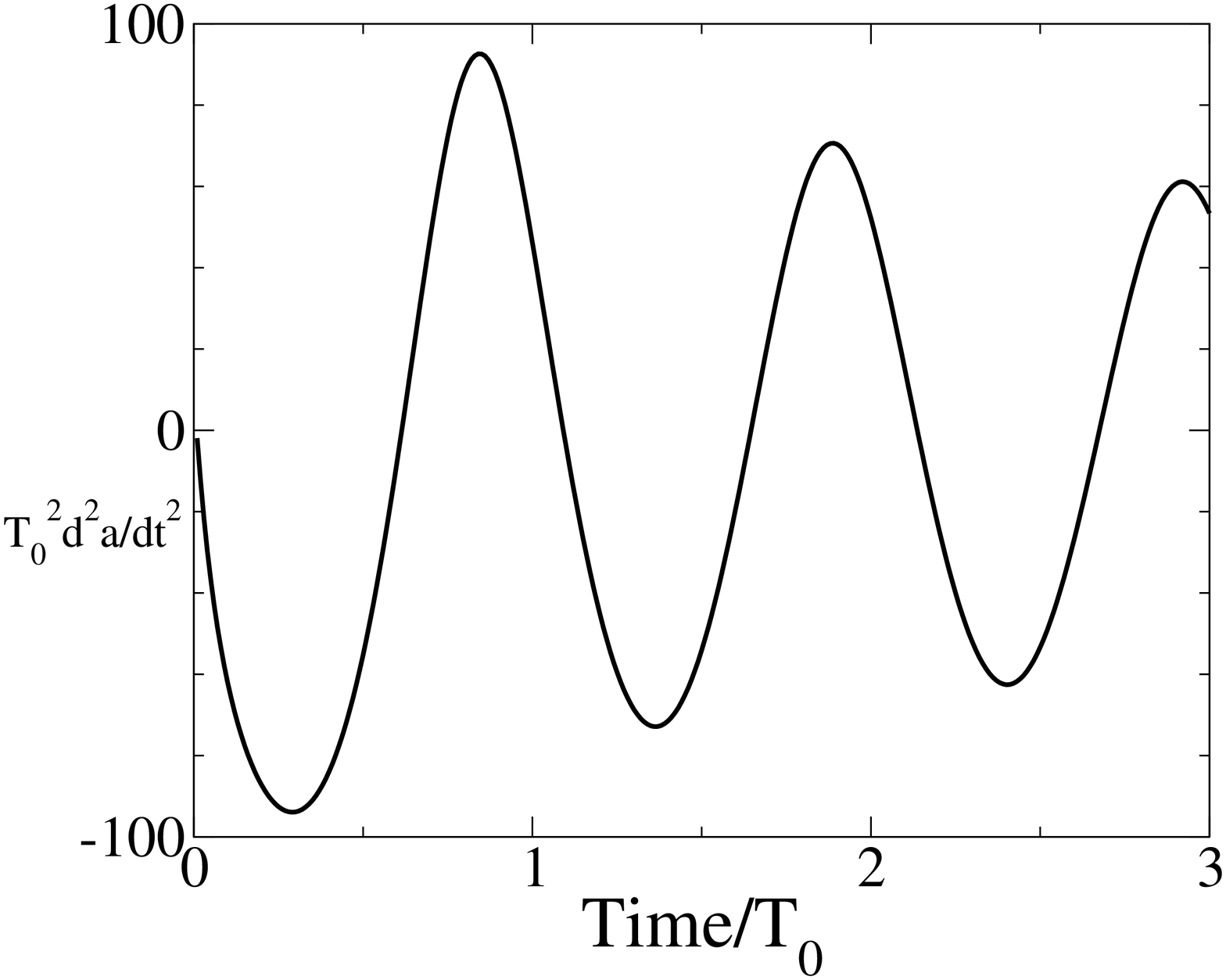} \\
\includegraphics[width=6cm]{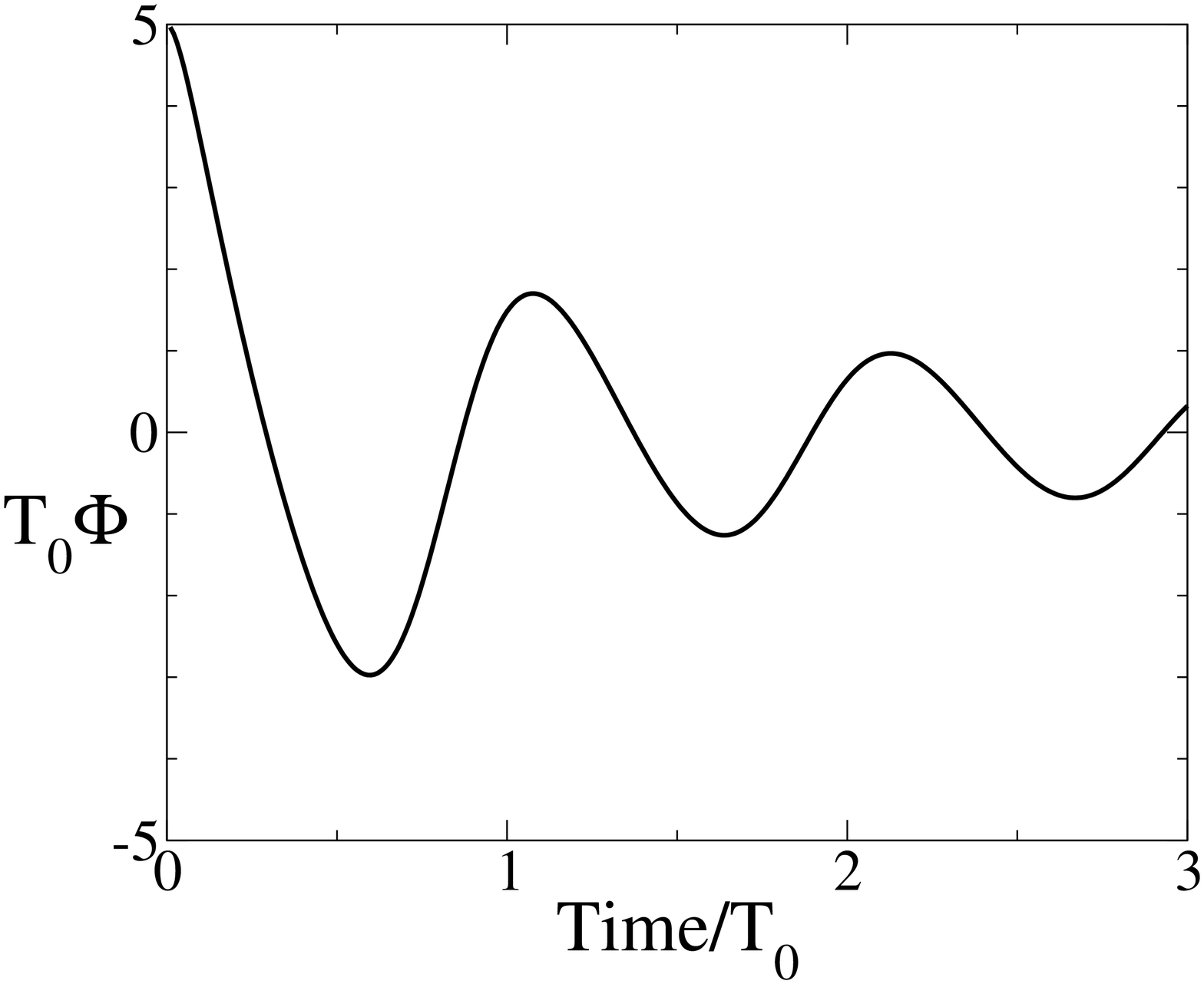}&
\includegraphics[width=6cm]{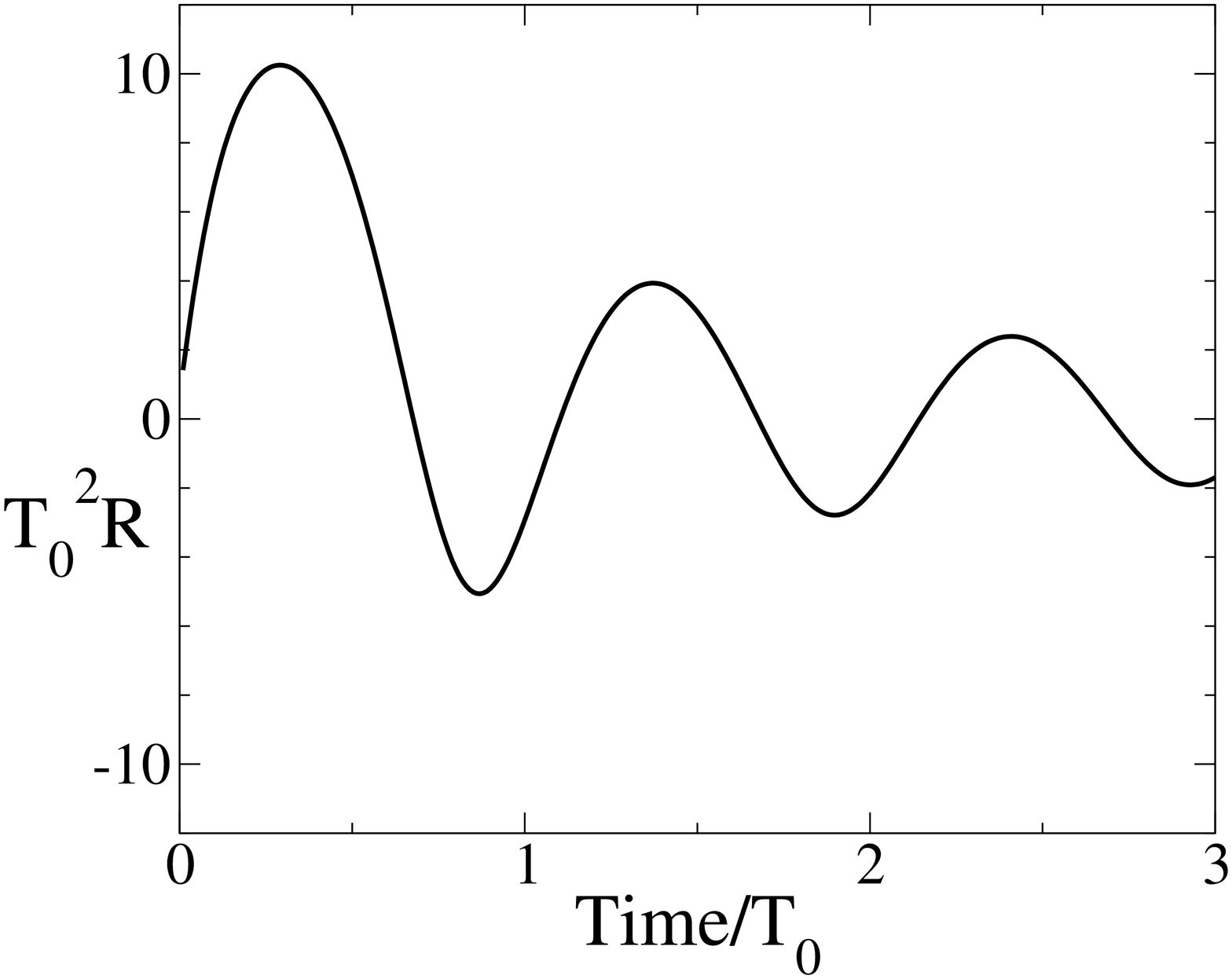}
\end{tabular}
\caption{Evolution of the Hubble function, the 2nd time derivative
of the expansion factor, the temporal component of the torsion, and
the affine scalar curvature as functions of time with the parameter
choice and the initial data in Case III.} \label{nowR}
\end{figure*}

%%%%%%%%%%%%%%%%%%%%%%%%%%%%%%%%%%%%%%%%
\subsection{Case I: constant $R$ case}
%%%%%%%%%%%%%%%%%%%%%%%%%%%%%%%%%%%%%%%%%%%%%%%%%%%%
In this case, the initial values of the fields are as follows:
$a(t_0)=10^5$, $H(t_0)=10^{-5}$, $\Phi(t_0)=2\times10^{-2}$,
$Y(t_0)=0$, and the parameters are taken to be $m=10^{-6}$, $\quad
b=8\times10^{-4}$. The result is shown in Fig.~\ref{conR}, where the
evolved values of $H$, $\ddot a$, $\Phi$, and $R$ are plotted in
different panels.

It is obvious in the bottom-right panel of Fig.~\ref{conR} that the
affine scalar curvature $R$ remains constant, $6m/b$. In other
words, $Y(t)=0$. The behavior of the torsion $\Phi$ can be
understood through Eq.~(\ref{dtphi}). $\Phi$ will increase (or
decrease) until its value balances the rhs of Eq.~(\ref{dtphi});
this mainly depends on the sign change of the term $3H\Phi$ provided
$H>0$ and ${\cal T}>0$. With the current initial choice in this
case, $\Phi$ decreases promptly until the balancing point is
reached, as seen in the bottom-left panel of Fig.~\ref{conR}.
However, $\Phi$ will not be a constant since the rhs of
Eq.~(\ref{dtphi}) still changes with time. The Hubble function $H$
will always decrease, as shown in the upper-left panel of
Fig.~\ref{conR}, since all the terms on the rhs of Eq.~(\ref{dtH})
are negative. Therefore, the acceleration of the expansion factor,
$\ddot a$, will be always negative, as seen in the upper-right panel
of Fig.~\ref{conR}. Obviously, the scenario described in this case
cannot explain the acceleration of the current Universe. However, it
could represent the very final stage of an oscillating universe
since the amplitude of the affine scalar curvature is damped to
$6m/b$, as described in the next two cases.
%%%%%%%%%%%%%%%%%%%%%%%%%%%%%%%%%%%%%%%%%%%%%%%%%%%%
\subsection{Case II: oscillating acceleration of $a$}
%%%%%%%%%%%%%%%%%%%%%%%%%%%%%%%%%%%%%%%%%%%%%%%%%%%%
For this case, we take the initial values of the field as follows:
$a(t_0)=10^8$, $H(t_0)=10^{-8}$, $\Phi(t_0)=10^{-4}$,
$Y(t_0)=-10^{-8}$ and the parameters are chosen to be $m=10^{-8}$,
$\quad b=8\times10^{-4}$. The results plotted in Fig.~\ref{oscR1}
show that $\ddot a$, $\Phi$, and $R$ are damped periodic.

Since $Y(t_0)$ is negative, $Y(t)$ remains negative. Thus
$R(t)<6m/b=7.5\times10^{-5}$ as shown in the bottom-right panel of
Fig.~\ref{oscR1}. Its period is about $T\approx3500$, different from
$2\pi\sqrt{b/2m}\approx1256$ by about a factor of 3. As explained in
the previous section, the difference comes from the choice of the
initial values. $\Phi$ has the same period with its magnitude close
to the maximal possible magnitude:
$\sqrt{9m/b}\approx1.6\times10^{-2}$. %in Eq.~(\ref{maxphi}).
The
most interesting part is the behavior of $\ddot a$, which is
periodic with the same period as $\Phi$ and $R$. As shown in the
top-right panel of Fig.~\ref{oscR1}, $\ddot a$ could be positive as
well as being negative and the pattern of its function is similar to
the pattern of $R$ with a minus sign. Therefore the behavior of $H$
is declining with a zig-zag pattern, as shown in the top-left panel
of Fig.~\ref{oscR1}.

In a broader viewpoint of the evolution of this system, $\ddot a$,
$\Phi$, and $R$ will be slowly damped and the zig-zag part of $H$
will be also smoothed after a long time. Eventually, the final stage
of this case will tend to be similar with Case I, i.e., $\ddot
a\leq0$, $R=\hbox{const}$. However, the important point of this case
is that the universe could oscillate due to dynamic torsion. In this
scenario  the present day acceleration of the Universe is not so
strange, $\ddot a$ is oscillating and it happens to be increasing at
this time. Furthermore, the period of the oscillation of a universe
 could be determined by the parameters and the initial values of
the fields in this model. This encouraged us to try to find
parameter values and initial conditions  which more nearly resemble
the current status of the Universe. Such a choice it will be shown
in the next case.
%%%%%%%%%%%%%%%%%%%%%%%%%%%%%%%%%%%%%%%%%%%%%%%%%%%%
\subsection{Case III: mimicking an accelerating universe at present}
%%%%%%%%%%%%%%%%%%%%%%%%%%%%%%%%%%%%%%%%%%%%%%%%%%%%
In this case, the initial values of the field are chosen as follows:
$a(t_0)=10$, $H(t_0)=4/T_0$, $\Phi(t_0)=5/T_0$, $Y(t_0)=-24/H^2_0$
and the parameters are taken to be $m=0.78$, $b=0.19T_0^2$,
to obtain a presently accelerating universe after a time on the
scale of the Hubble time.

In the top-left panel of Fig.~\ref{nowR}, the Hubble function $H$
evolution is fast-damped and oscillating, with the magnitude being
on the order of $1/T_0$ at a time near $T_0$ (the scale of the
Hubble time). Moreover, $H$ is increasing at that time epoch. We can
see this more clearly from the top-right panel of Fig.~\ref{nowR},
where $\ddot a$ is plotted. It is obvious that $\ddot a$ is damped
and oscillating during the evolution and is positive around
$t\approx T_0$. In this setting, $R$, shown in the bottom-right
panel of Fig.~\ref{nowR}, is smaller than $6m/b=24/T_0^2$, as stated
in Eq.~(\ref{Rrange}). Although the initial values of $\Phi$ and $H$
are similar in this case, the magnitude of $H$ is decreasing faster
than that of $\Phi$. This verifies that we can usually assume
$\Phi\gg H$. Here the behavior of all the fields at the very
beginning of the time should not be taken too seriously since in the
early universe our matter-dominated era assumption breaks down. We
hope to discuss a radiation-dominated universe with this model in
the future.

Although we cannot know if the initial data and parameter values
chosen here are realistic without a further investigation, this case
does demonstrate the possibility of explaining the accelerating
expansion of the universe with the dynamic scalar torsion PGT model.
It should be noted that the accelerating expansion simply comes
about because of the current phase of the oscillation of the
universe. Such an oscillation mechanism may offer a good answer to
the related well known coincidence problem. Hence it will be very
interesting to see if this model can be consistent with  up-to-date
data from  astronomical observations.
%%%%%%%%%%%%%%%%%%%%%%%%%%%%%%%%%%%%%%%
\section{Discussion}
%%%%%%%%%%%%%%%%%%%%%%%%%%%%%%%%%%%%%%%%
In this work we introduce into the evolution of a universe without a
cosmological constant a certain dynamical PGT scalar torsion mode
taken from our earlier work.\cite{YHNJ99} From the assumption of the
homogeneity and isotropy of the universe, only the temporal
component of the torsion $\Phi$ will survive and affect the
evolution of the universe at late times. With the field equations
(\ref{dta}--\ref{dtR}), we analyzed analytically and numerically the
evolution of the system. We found that there is a critical value of
the  affine curvature $R$, i.e., $R=6m/b$, which divides the system
into two different behaviors. If the initial value $R(t_0)>6m/b$,
then $R(t)$ is always greater than $6m/b$ and the amplitudes of all
the variables may grow unboundedly. If $R(t_0)<6m/b$, then $R(t)$
will be always less than $6m/b$ and $\ddot a$, $\Phi$, $R$, tend to
have a damped periodic behavior with the same period. With certain
choices of the parameters of $m$ and $b$, and of the initial data of
$a$, $H$, $\Phi$, and $R$, like Case III in the previous section,
this model can describe an oscillating universe with an accelerating
expansion at the present time. If we consider instead the spacetimes
as Riemannian, by absorbing the contribution of the torsion of this
model into the stress-energy tensor on the rhs of the Einstein's
equation, then this contribution will act like an {\it exotic} fluid
with its mass density $\rho_{\rm T}$ and pressure $p_{\rm T}$
varying with respect of time. Moreover, it presently has a negative
pressure and consequently a negative parameter in the equation of
state, i.e., $\omega_{\rm T}$, which drives the universe into
accelerating expansion.

Before we can give an adequate discussion of the viability of this
model as an explanation of the accelerating universe, we should
check whether this model can survive under the constraints of the
theoretical and experimental tests. There have been numerous
investigations on the existence of torsion since this geometric
quantity entered the realm of gravity (see
[\refcite{HaRT02,ShIL02,AHPJ04}], and the references therein.) As
mentioned above, this model has not only passed the important
classical tests
--- ``no-ghosts" and ``no-tachyons" ---
it is one of the two scalar torsion modes, the only PGT cases which
are known to have a well posed initial value problem\cite{YHNJ99}
and which may well be the only viable dynamic PGT torsion modes that
can evade the non-linear constraint problems.   There have been a
lot of laboratory tests in search of torsion.\cite{CTNW93,NiWT96}
The main idea among these experiments is the spin interaction
between matter and torsion. The cosmological tests on torsion
investigate the effect of torsion-induced spin flips of neutrinos in
the early Universe which could alter the helium abundance and have
other effects on the early nucleosynthesis.\cite{CILS99,BruM99}
However Dirac fermions interact only with the totally antisymmetric
pseudo-scalar part of the torsion. Thus these tests can only
consider the pseudo-scalar-mode/axial-vector torsion, not the scalar
mode torsion used in our model. The type of torsion used in our
model does not interact directly with any known matter. Thus, these
tests cannot really give a serious constraint on the amplitude of
our scalar-mode torsion.

Among the models in which torsion is applied to the cosmological
problem, Capozziello {\it et al} \cite{CSet03,CaCT03} have done a
serious study on replacing the role of the cosmological constant in
the accelerating Universe. With a totally antisymmetric torsion
without dynamical evolution, their model is consistent with the
observational data by tuning the amount of the torsion density,
although this model cannot solve the coincidence problem. On the
other hand, the oscillating universe models with a designed
mechanism: an oscillating potential, an oscillating parameter of the
equation of state, etc.,\cite{DoKS00,RSPC03,FLPZ06} aim to solve the
coincidence problem. Here we found that our model takes some virtues
from both kind of models, i.e., our model is capable of solving the
coincidence problem of an accelerating universe with a dynamical
scalar-mode torsion, which is {\it naturally} obtained from the
Geometry of the Riemann-Cartan spacetimes, instead of an exotic
scalar field or a designed mechanism.

In this work we proposed the scalar torsion mode of the PGT  as a
viable model for explaining the current status of the Universe. The
source of the torsion could come from the huge density of the
particles with the aligned spins in the early universe. This scalar
mode of torsion could be considered as a ``phantom" field, at least
in the matter-dominated epoch, since it will not interact directly
with matter; it only interacts indirectly via gravitation. Then the
dynamics of the scalar torsion mode could drive the Universe in an
oscillating fashion with an accelerating expansion at present.
However, there are also some points which need to be studied in much
more detail before this model can more closely conform to reality.
The model in Case III of the previous section, suggests that the
mass parameter of the torsion, $m$, might be close to $a_0$, and the
parameter for the ``kinetic'' energy density of the torsion, $b$,
may need to be as huge as $T_0^2$ to achieve an accelerating
universe. The restricted window of the parameter choices which
allows a behavior like that of our universe might render the model
less favored, even though the matter in the universe is not able to
directly interact with the torsion. Meanwhile, the required choice
of initial data and the values of the parameters may make this model
unsuited to solving the fine-tuning problem. However, these dark
sides will not be able to diminish the possibility of the scalar
mode of the torsion in this model playing a magnificent role in the
the evolution of the Universe. Moreover, we only used one of the
viable modes of torsion in PGT, and the argument may be more general
if it can be extended to all the viable PGT torsion modes. Further
studies on model building, a comparison between the observational
data and the predictions of this model, and parameter/initial data
determination are currently under investigation.\cite{ShNY07}
%%%%%%%%%%%%%%%%%%%%%%%%%%%%%%%%%%%%%%%%
\section*{Acknowledgments}
%%%%%%%%%%%%%%%%%%%%%%%%%%%%%%%%%%%%%%%%
The authors are grateful to Chopin Soo, Rue-Ron Hsu, and Kuo-Feng
Shie for their helpful suggestions and discussions. This work was
supported in part by the National Science Council of the R.O.C.
(Taiwan) under grant Nos.~NSC94-2112-M-006-014,
NSC95-2112-M-006-017-MY2 and NSC 95-2119-M008-027.  Some of the
calculations were performed at the National Center for
High-performance Computing in Taiwan.

\end{document}